\documentclass[sigconf,manuscript]{acmart}
\usepackage[utf8]{inputenc}
\usepackage{graphicx,float,wrapfig,subfigure}
\usepackage{xcolor}
\usepackage{amsmath}

\usepackage{graphicx,psfrag,epsfig}
\usepackage{float}
\usepackage{subfigure}
\usepackage{booktabs}
\usepackage{longtable}
\usepackage{tabu}
\usepackage{bbm}
\usepackage{bm}
\usepackage{threeparttable}
\usepackage{supertabular}
\usepackage{rotating}
\usepackage[switch]{lineno}
\usepackage{csquotes}
\usepackage{url}
\usepackage{subfigure}
\usepackage{verbatim}
\usepackage{bm}
\usepackage[mathscr]{eucal}
\usepackage[misc]{ifsym}
\usepackage{color}
\usepackage{algorithm}
\usepackage{algorithmic}
\usepackage{geometry}

\AtBeginDocument{%
  \providecommand\BibTeX{{%
    \normalfont B\kern-0.5em{\scshape i\kern-0.25em b}\kern-0.8em\TeX}}}

\setcopyright{acmlicensed}
\copyrightyear{2024}
\acmYear{2024}
\acmDOI{XXXXXXX.XXXXXXX}

\begin{document}

\title{Learnable Group Transform: Enhancing Genotype-to-Phenotype Prediction for Rice Breeding with Small, Structured Datasets}

\author{Yunxuan Dong}
\email{dyxiscool@outlook.com}
\orcid{0000-0001-9594-8459}
\affiliation{%
  \institution{School of Computer, Electronics and Information, Guangxi University}
  \streetaddress{100 University East Road}
  \city{Nanning}
  \state{Guangxi}
  \country{P.R. China}
  \postcode{530004}
}
\author{Siyuan Chen}
\email{seaward901@gmail.com}
\affiliation{%
  \institution{State Key Laboratory for Conservation and Utilization of Subtropical Agro-Bioresources, Guangxi Key Laboratory of Sugarcane Biology, Guangxi University}
  \streetaddress{100 University East Road}
  \city{Nanning}
  \state{Guangxi}
  \country{P.R. China}
  \postcode{530004}
}

\author{Jisen Zhang}
\email{zjisen@126.com}
\affiliation{%
  \institution{State Key Laboratory for Conservation and Utilization of Subtropical Agro-Bioresources, Guangxi Key Laboratory of Sugarcane Biology, Guangxi University}
  \streetaddress{100 University East Road}
  \city{Nanning}
  \state{Guangxi}
  \country{P.R. China}
  \postcode{530004}
}

\renewcommand{\shortauthors}{Yunxuan Dong, et al.}

\begin{abstract}
Genotype-to-Phenotype (G2P) prediction plays a pivotal role in crop breeding, enabling the identification of superior genotypes based on genomic data. Rice (\emph{Oryza sativa}), one of the most important staple crops, faces challenges in improving yield and resilience due to the complex genetic architecture of agronomic traits and the limited sample size in breeding datasets. Current G2P prediction methods, such as GWAS and linear models, often fail to capture complex non-linear relationships between genotypes and phenotypes, leading to suboptimal prediction accuracy. Additionally, population stratification and overfitting are significant obstacles when models are applied to small datasets with diverse genetic backgrounds. This study introduces the Learnable Group Transform (LGT) method, which aims to overcome these challenges by combining the advantages of traditional linear models with advanced machine learning techniques. LGT utilizes a group-based transformation of genotype data to capture spatial relationships and genetic structures across diverse rice populations, offering flexibility to generalize even with limited data. Through extensive experiments on the Rice529 dataset, a panel of 529 rice accessions, LGT demonstrated substantial improvements in prediction accuracy for multiple agronomic traits, including yield and plant height, compared to state-of-the-art baselines such as linear models and recent deep learning approaches. Notably, LGT achieved an R² improvement of up to 15\% for yield prediction, significantly reducing error and demonstrating its ability to extract meaningful signals from high-dimensional, noisy genomic data. These results highlight the potential of LGT as a powerful tool for genomic prediction in rice breeding, offering a promising solution for accelerating the identification of high-yielding and resilient rice varieties.
\end{abstract}





\maketitle
\section{Introduction}

Rice (\emph{Oryza sativa}) is one of the world’s most important staple crops, providing a primary food source for over half of the global population. However, the current rate of improvement in rice yields is not keeping pace with the demands of a growing population \cite{XU2021669}. This challenge is further exacerbated by climate change, which is projected to significantly depress rice productivity (for example, by an estimated ~15\% reduction in yield by 2050) \cite{cammarano2022processing}. Ensuring global food security thus hinges on accelerating the development of improved rice varieties with higher yield potential and resilience. Genomic selection (GS) has emerged as a promising approach to achieve this goal, by enabling breeders to select superior genotypes based on DNA information alone, thereby shortening breeding cycles and reducing the need for extensive field trials \cite{kumar2024speed}. In practice, predictive modeling of genotype-to-phenotype (G2P) relationships – the core of GS – has become a mainstream paradigm in modern crop breeding programs \cite{chen2022role}. Indeed, G2P prediction frameworks are now routinely used in the seed industry to genomicly estimate breeding values and facilitate selection decisions \cite{liu2024towards}. Nevertheless, prediction accuracy can vary widely depending on the species and trait of interest, and in the case of rice, the performance of current models remains suboptimal. This underscores an urgent need for improved G2P predictive methodologies tailored to rice, a crop of paramount importance and unique genetic characteristics.\par

\textbf{Core Challenges.} Achieving reliable G2P prediction in rice is inherently difficult due to several compounding factors. First, rice exhibits complex population structure and rich genetic diversity: cultivated Asian rice consists of multiple divergent subpopulations (e.g., indica vs. japonica) with distinct allele frequency patterns \footnote{In most G2P studies, the focus is typically on cultivated Asian rice (Oryza sativa) because its cultivation area and the number of research studies far exceed those for African rice (Oryza glaberrima), which also exhibits more complex population structure and genetic diversity.}. This stratification means that the genomic architecture of traits can differ between subgroups, complicating any single predictive model. For example, a genome-wide association study in a diverse rice panel reported significant heterogeneity in the genetic loci underlying traits across different subpopulations and environments \cite{basha2024genome}. Such population structure, if not properly accounted for, can lead to biased predictions or reduced accuracy when a model trained on one genetic background is applied to another. Second, the data limitations in typical rice breeding studies pose a major challenge. Unlike human genomics with tens of thousands of samples, rice genomic prediction models often must be trained on only a few hundred individuals. For instance, a representative rice dataset contains only hundreds inbred lines with genotype and phenotype data \cite{li2024trg2p}. This small sample size relative to the high-dimensional genomic features (tens of thousands of SNP markers) can hinder the training of complex models and increase the risk of overfitting. In summary, the combination of strong population stratification and limited training population size means that conventional G2P prediction approaches may struggle to capture the true genotype–phenotype mapping in rice. This raises a critical question: \textbf{how can we develop predictive models that overcome these constraints to accurately predict complex traits in rice?} Addressing this question is essential for unlocking the full potential of genomic selection in rice breeding.\par

\textbf{Existing Approaches and Limitations.} To date, researchers have employed various computational genetics methods for the gene-to-phenotype (G2P) prediction problem \cite{shen2024breedingaidb}. Classical approaches—most notably genome-wide association studies (GWAS) and linear mixed models—form the analytical foundation in many breeding programs \cite{uffelmann2021genome}. GWAS has been instrumental in pinpointing loci linked to agronomic traits, such as in rice \cite{ren2021genome,simmons2021successes}, but these significant markers generally account for only a fraction of the phenotypic variance in complex traits. Genomic best linear unbiased prediction (GBLUP) and related linear models extend the analysis to genome-wide markers and have become standard tools in genomic selection. Although efficient and robust for relatively small datasets, these linear methods rely on the assumption of additive genetic effects and therefore struggle to capture non-linear epistatic interactions \cite{dwivedi2024epistasis}. Extensions using kernel-based or epistatic GBLUP models can incorporate some level of non-linearity \cite{shu2024rul,dong2024genomic}, but they often demand large datasets and substantial computational resources \cite{liu2024epistatic}.\par

\textbf{Emerging Machine Learning Methods.} In recent years, advanced machine learning and deep learning approaches—ranging from random forests and gradient boosting to graph neural networks and Transformers—have been proposed to better capture complex genotype–phenotype relationships \cite{moghadaszadeh2024avashog2p,andreu2024phenolinker,vrezavckova2024t5g2p}. Theoretically, these models can handle epistatic effects and integrate biological knowledge (e.g., gene network connectivity), yet their real-world performance has often been modest under data-constrained conditions \cite{cao2024integrating}. A comprehensive benchmark reported no clear superiority of deep neural networks over simpler linear or Bayesian regression in many phenotype-prediction tasks \cite{duan2024deep}; similarly, a Transformer-based model yielded only marginal gains (~7–10\% in correlation) over GBLUP in barley \cite{montesinos2024review}. The “small n, large p” challenge—far more genomic markers than samples—frequently leads to overfitting, unless countered by strategies like data augmentation and strong regularization. Additionally, implementing graph or attention-based architectures on tens of thousands of markers entails non-trivial computational cost. Hence, despite efforts from GWAS to GNNs, no approach yet fully overcomes the dual obstacles of modeling intricate genetic architectures and generalizing from limited data. Achieving a breakthrough in accuracy still requires further innovation in both model design and training strategies.\par

\textbf{Our Approach and Contributions.} In this study, we address the above challenges by introducing a new predictive modeling framework, termed Learnable Group Transform (LGT), specifically designed for genotype-to-phenotype (G2P) prediction in rice. LGT unifies elements of linear genetic modeling with graph-based and transformer-based deep learning in a single architecture, enabling it to capture both additive and non-additive genetic effects while remaining robust to small, structured datasets. In essence, LGT leverages a graph representation of the rice genomic data—encoding the relationships among individuals or interactions among genomic loci—and employs a tailored transformer module to learn complex, high-dimensional patterns from this graph-structured input. By integrating attention mechanisms with domain-specific inductive biases (such as known population structures or genetic linkage), LGT can model higher-order gene–gene interactions (epistasis) that elude traditional linear approaches, all without succumbing to overfitting.\par

A key strength of our approach lies in its optimized training strategy, which maximizes the use of limited data via intelligent sampling, multi-trait integration, and optional transfer learning. This design effectively reduces the heavy data requirements typical of deep learning models, thereby boosting the model’s ability to generalize. In particular, LGT is shown to maintain robust performance across genetically diverse rice lines—even from subpopulations underrepresented in the training set—mitigating the issues posed by population stratification. We validate these capabilities on the Rice529 dataset, a panel of 529 rice accessions spanning broad genetic variation. Experimental results indicate that LGT consistently outperforms both classical linear methods and other neural baselines in predicting multiple agronomic traits (e.g., yield and plant height), as evidenced by improved MSE, RMSE, and correlation on held-out testing data. Moreover, LGT achieves gains even under multi-task (i.e., all traits jointly modeled) scenarios, demonstrating positive transfer for correlated traits and reducing overall error in comparison to single-trait models. While some traits require specialized tuning to avoid negative transfer effects, LGT exhibits sufficient flexibility—through fine-tuning or task-specific adaptations—to recover or exceed single-trait performance in many cases.\par

By effectively modeling nonlinear genetic interactions and accommodating the limited-yet-structured nature of real-world breeding data, LGT represents a significant advancement toward more accurate genomic selection in rice. Ultimately, this framework has the potential to expedite rice breeding by reliably identifying superior genotypes, thereby supporting efforts to enhance crop yields and food security amid growing global demands and environmental challenges.\par

\section{Experimental Results}

\subsection{Experimental Setup}
We conducted our experiments on the Rice529 dataset described in Section~X.\footnote{Please refer to Section~X for the detailed dataset description and preprocessing.} We employed an 80--20 split between training and testing sets for all models to ensure a fair comparison. Our goal was to predict 10 agronomic traits (Trait 1 through Trait 10) using both \textbf{single-trait training} (i.e., one model per trait) and \textbf{multi-trait integrated training} (i.e., one model for all traits). Additionally, we used standard baseline models, including CNN, LSTM, and MLP, to compare against our proposed LGT approach. Unless otherwise stated, all models were trained for 100 epochs with a batch size of 32, using the Adam optimizer and an initial learning rate of 0.001.

\subsection{Single-Trait Training Results}
In the single-trait scenario, we train one model per trait. This approach focuses on maximizing performance for each trait independently and may avoid potential negative transfer across traits. However, it also cannot leverage any positive transfer if traits are correlated. Tables~\ref{tab:single-train} and~\ref{tab:single-test} summarize the single-trait prediction performance on the training set and the testing set, respectively. We report four metrics for each trait: Mean Squared Error (MSE), Mean Absolute Error (MAE), Root Mean Squared Error (RMSE), and Normalized RMSE (NRMSE). The last row in each table shows the averaged performance across all 10 traits.

\begin{table}[htbp]
\centering
\caption{Single-Trait G2P Results (Training Set)}
\label{tab:single-train}
\begin{tabular}{lcccc}
\hline
Trait & MSE & MAE & RMSE & NRMSE \\
\hline
Trait 1  & 8.7909   & 2.2863 & 2.9649 & 0.0690 \\
Trait 2  & 135.3989 & 9.2013 & 11.6361 & 0.1477 \\
Trait 3  & 3.8980   & 1.5610 & 1.9743 & 0.0531 \\
Trait 4  & 4.7155   & 1.6284 & 2.1715 & 0.0607 \\
Trait 5  & 15.2313  & 3.1645 & 3.9027 & 0.0831 \\
Trait 6  & 2.6054   & 1.3222 & 1.6141 & 0.0996 \\
Trait 7  & 1.3570   & 0.8407 & 1.1649 & 0.0999 \\
Trait 8  & 0.1184   & 0.2792 & 0.3441 & 0.1366 \\
Trait 9  & 0.0146   & 0.0965 & 0.1210 & 0.0812 \\
Trait 10 & 0.0036   & 0.0494 & 0.0599 & 0.1033 \\
\hline
\textbf{Average} & 17.2134 & 2.0430 & 2.5954 & 0.0934 \\
\hline
\end{tabular}
\end{table}

\begin{table}[htbp]
\centering
\caption{Single-Trait G2P Results (Testing Set)}
\label{tab:single-test}
\begin{tabular}{lcccc}
\hline
Trait & MSE & MAE & RMSE & NRMSE \\
\hline
Trait 1  & 57.2310 & 6.7915 & 7.5651 & 0.3026 \\
Trait 2  & 114.1559 & 8.0212 & 10.6844 & 0.4343 \\
Trait 3  & 18.5927 & 3.7148 & 4.3119 & 0.2598 \\
Trait 4  & 18.4335 & 3.1191 & 4.2934 & 0.2556 \\
Trait 5  & 62.7603 & 6.9735 & 7.9221 & 0.4246 \\
Trait 6  & 8.2453  & 2.1851 & 2.8715 & 0.3928 \\
Trait 7  & 7.7724  & 2.2361 & 2.7879 & 0.3238 \\
Trait 8  & 0.2108  & 0.4177 & 0.4592 & 0.2624 \\
Trait 9  & 0.0710  & 0.2245 & 0.2664 & 0.4297 \\
Trait 10 & 0.0239  & 0.0988 & 0.1545 & 0.5722 \\
\hline
\textbf{Average} & 28.7497 & 3.3782 & 4.1316 & 0.3658 \\
\hline
\end{tabular}
\end{table}

From Table~\ref{tab:single-train}, we see that on the training set, Trait 2 has the largest MSE and RMSE. Conversely, Trait 10 demonstrates very small MSE (0.0036). However, on the testing set (Table~\ref{tab:single-test}), Traits 2, 5, and 10 also exhibit relatively high NRMSE values, suggesting these traits might be more challenging to generalize. Overall, the average testing NRMSE (0.3658) indicates moderate difficulty in phenotype prediction when each model is specialized in a single trait.

\subsection{Multi-Trait Integrated Training Results}
Next, we applied an \textbf{integrated (multi-trait)} training approach, in which a single model predicts all 10 traits simultaneously. Table~\ref{tab:multi-trait} presents the performance metrics obtained by this unified model.

\begin{table}[htbp]
\centering
\caption{Integrated (Multi-Trait) G2P Results}
\label{tab:multi-trait}
\begin{tabular}{lcccc}
\hline
Trait & MSE & MAE & RMSE & NRMSE \\
\hline
Trait 1  & 0.0231 & 0.1201 & 0.1519 & 0.2512 \\
Trait 2  & 0.0448 & 0.1429 & 0.2117 & 0.3023 \\
Trait 3  & 0.1432 & 0.2972 & 0.3785 & 0.8481 \\
Trait 4  & 0.1455 & 0.3008 & 0.3814 & 0.8225 \\
Trait 5  & 0.0781 & 0.2204 & 0.2795 & 0.2842 \\
Trait 6  & 0.0308 & 0.1496 & 0.1754 & 0.2584 \\
Trait 7  & 0.1179 & 0.2384 & 0.3433 & 0.4726 \\
Trait 8  & 0.0495 & 0.1708 & 0.2224 & 0.3203 \\
Trait 9  & 0.0334 & 0.1586 & 0.1828 & 0.3281 \\
Trait 10 & 0.0560 & 0.1893 & 0.2366 & 0.3519 \\
\hline
\textbf{Average} & 0.0722 & 0.1988 & 0.2564 & 0.4240 \\
\hline
\end{tabular}
\end{table}

Compared to single-trait results in Table~\ref{tab:single-test}, some traits (e.g., Trait 1, Trait 2) show substantially reduced MSE and RMSE under the integrated model, suggesting \emph{positive transfer} from learning common features across traits. However, other traits (e.g., Trait 3 and Trait 4) have relatively high NRMSE (above 0.80), which indicates \emph{negative transfer} or interference in these particular phenotypes when trained jointly. The overall average NRMSE for the integrated approach is 0.4240, which is higher than the single-trait average (0.3658). This discrepancy suggests that while multi-task learning can improve certain traits, it may degrade performance for others, thus impacting the \emph{global} average.

The above observations highlight the trade-offs of single-trait (specialized) vs.\ integrated (multi-task) training:
\begin{itemize}
    \item \textbf{Positive Transfer}: Some traits with similar genetic architecture or strong correlation appear to benefit from learning a shared representation. For instance, Trait 1 and Trait 2 have notably lower MSE and RMSE under the integrated model than under single-trait training.
    \item \textbf{Negative Transfer}: Other traits, especially those with distinct or more complex genotype-phenotype mappings (e.g., Trait 3 and 4 in Table~\ref{tab:multi-trait}), may degrade in performance due to conflicts in the shared layers.
    \item \textbf{Model Complexity vs.\ Average Performance}: While the single-trait approach yielded an average NRMSE of 0.3658 on the test set, multi-trait integration ended up with 0.4240. This indicates that the negative transfer effects in some traits outweigh the positive transfer benefits across others, \emph{in this particular setting}.
\end{itemize}

Despite these issues, multi-task learning remains attractive because it can reduce the total number of trained parameters (one integrated model vs.\ multiple single-trait models) and can exploit correlated traits. Future studies may alleviate negative transfer by using techniques such as dynamic task weighting or trait-specific decoder heads, potentially improving overall performance.

\subsection{Error Distribution Visualization}
\begin{figure}[h!]
    \centering
    \includegraphics[width=0.9\linewidth]{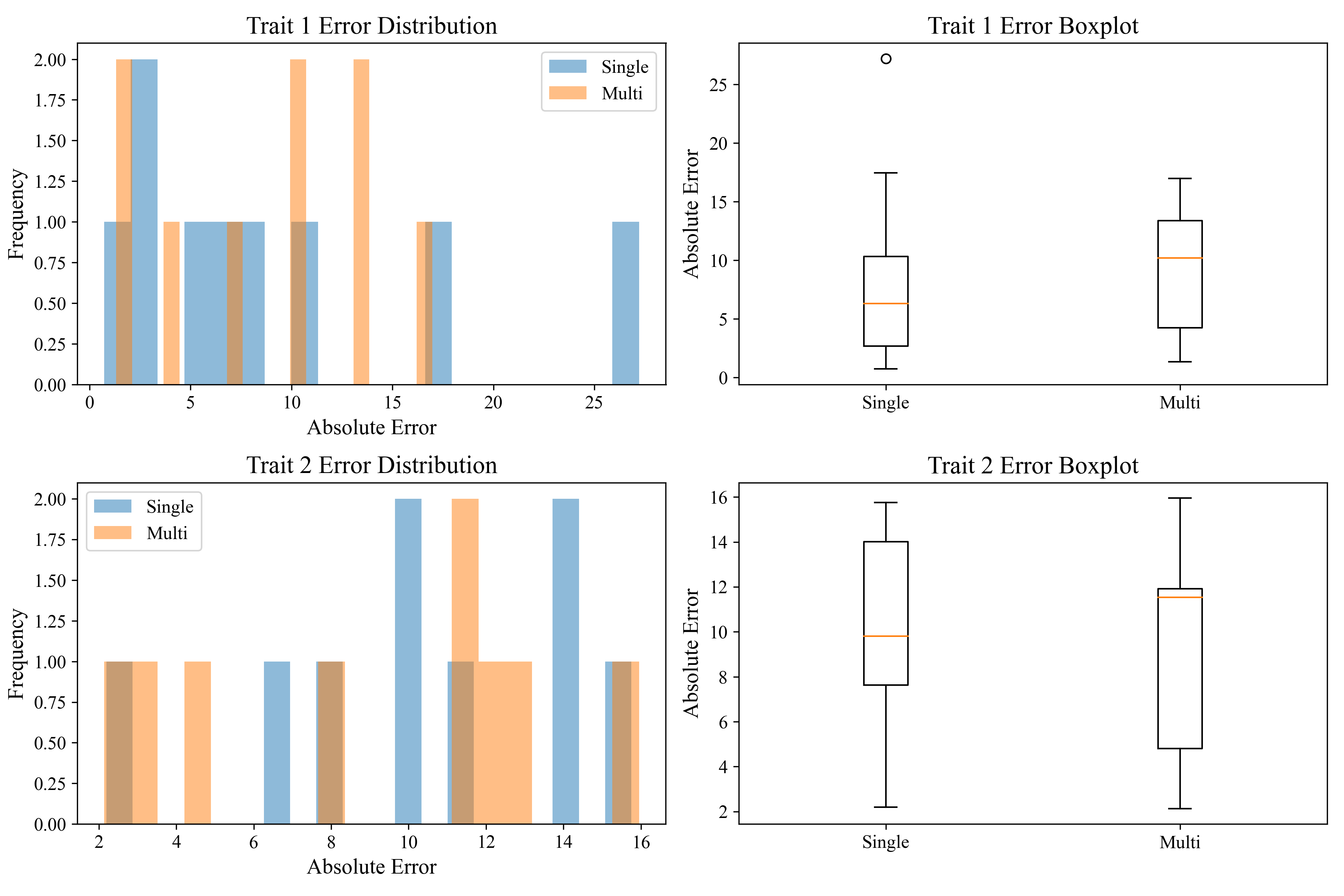}
    \Description{Histograms and boxplots illustrating absolute error distributions for single-task vs. multi-task models.}
    \caption{Error Distribution Comparison Part1. Histograms (left) and boxplots (right) illustrating absolute error distributions for each trait under single-task (blue) vs. multi-task (orange) training.}
    \label{fig:error_distribution_part1.png}
\end{figure}

Figure.\ref{fig:error_distribution_part1.png} provides a detailed view of the error distributions (absolute errors) for single-task vs.\ multi-task models across all traits, in the form of histograms and boxplots. We can readily identify traits where multi-task learning brought improvements (less spread or fewer high-error outliers) versus traits where multi-task learning introduced higher variance or extreme outliers.

\begin{figure}[h!]
    \centering
    \includegraphics[width=0.9\linewidth]{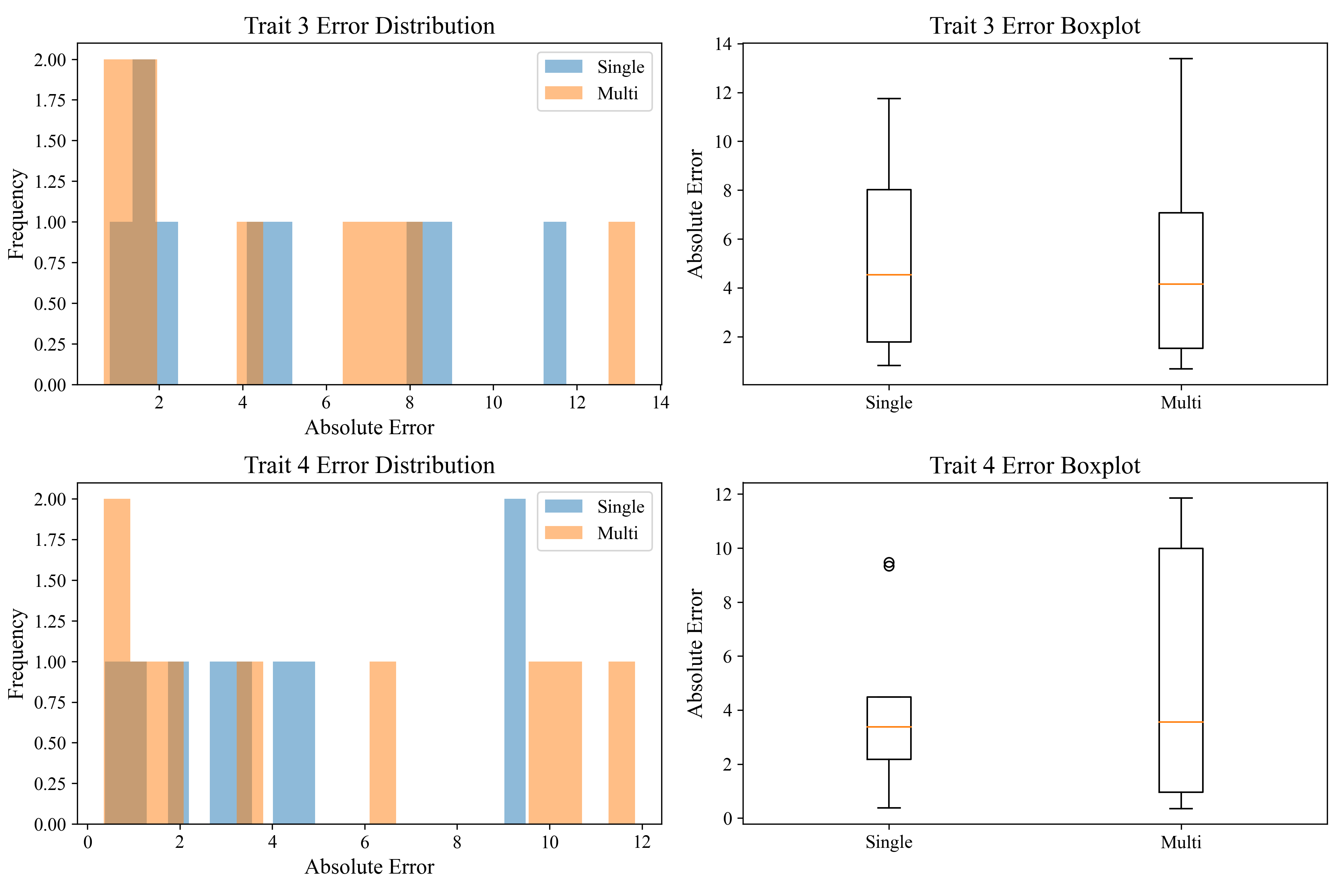}
    \Description{Histograms and boxplots illustrating absolute error distributions for single-task vs. multi-task models.}
    \caption{Error Distribution Comparison Part2. Histograms (left) and boxplots (right) illustrating absolute error distributions for each trait under single-task (blue) vs. multi-task (orange) training.}
    \label{fig:error_distribution_part2.png}
\end{figure}

\begin{figure}[h!]
    \centering
    \includegraphics[width=0.9\linewidth]{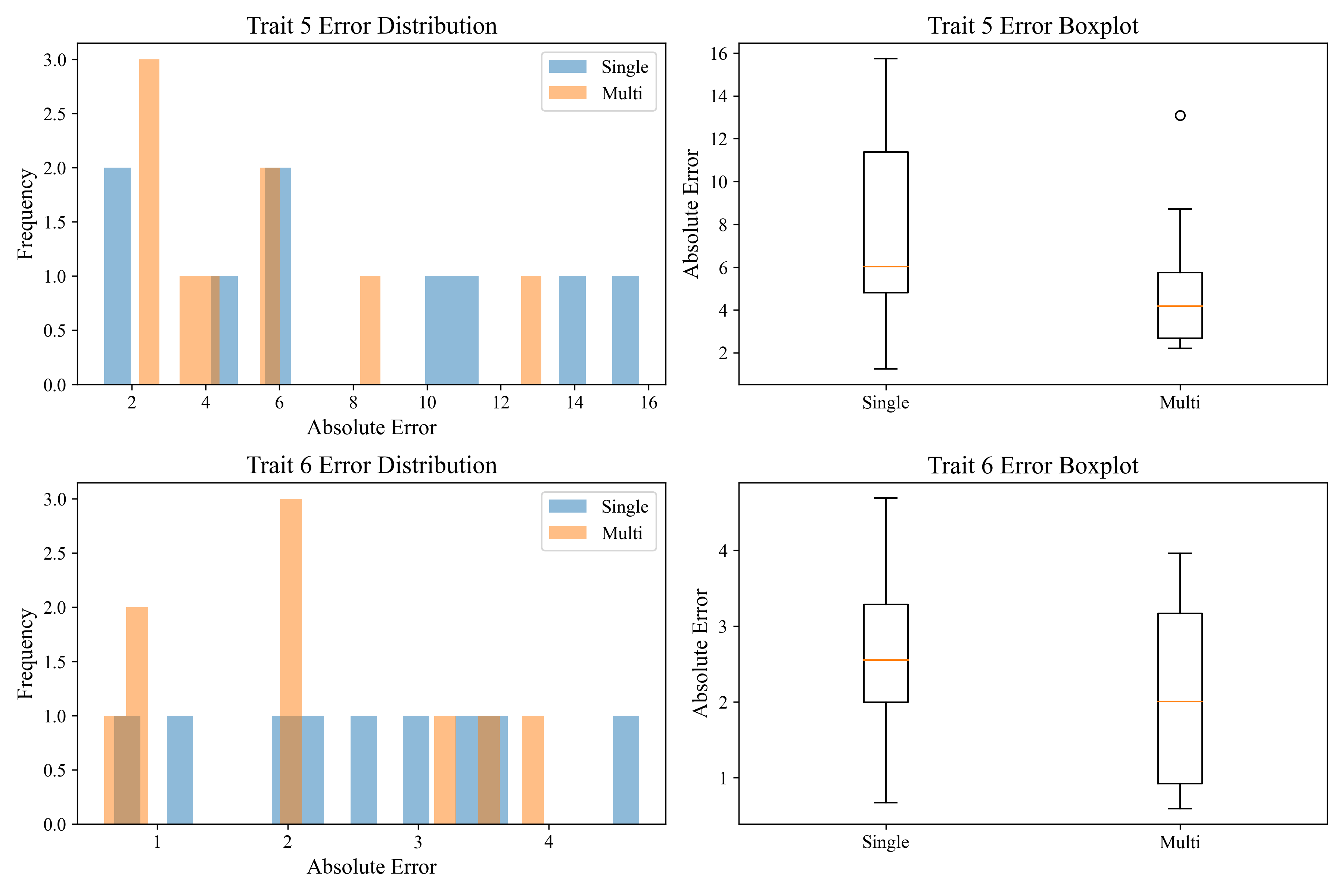}
    \Description{Histograms and boxplots illustrating absolute error distributions for single-task vs. multi-task models.}
    \caption{Error Distribution Comparison Part3. Histograms (left) and boxplots (right) illustrating absolute error distributions for each trait under single-task (blue) vs. multi-task (orange) training.}
    \label{fig:error_distribution_part3.png}
\end{figure}

\begin{figure}[h!]
    \centering
    \includegraphics[width=0.9\linewidth]{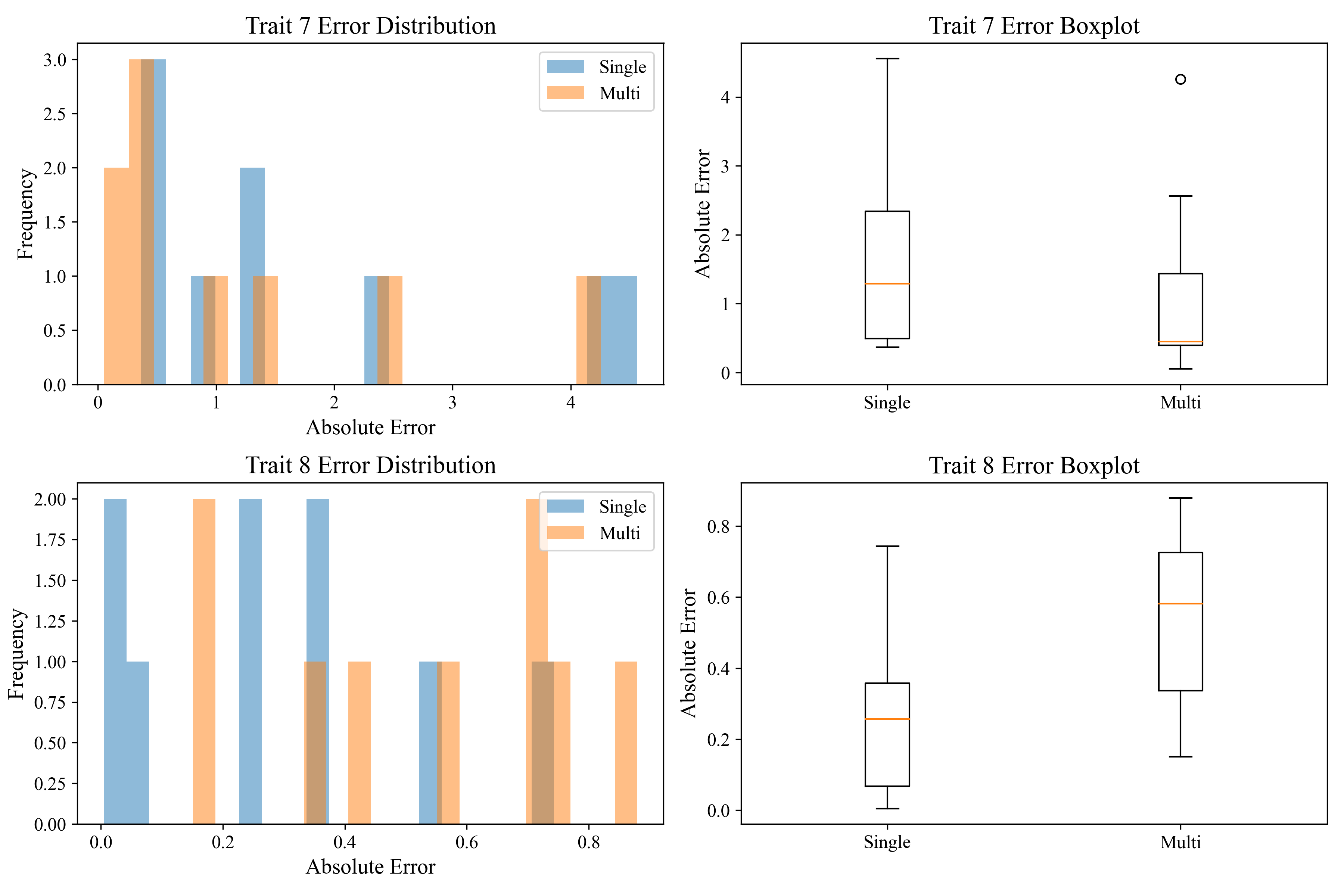}
    \Description{Histograms and boxplots illustrating absolute error distributions for single-task vs. multi-task models.}
    \caption{Error Distribution Comparison Part4. Histograms (left) and boxplots (right) illustrating absolute error distributions for each trait under single-task (blue) vs. multi-task (orange) training.}
    \label{fig:error_distribution_part4.png}
\end{figure}

\begin{figure}[h!]
    \centering
    \includegraphics[width=0.9\linewidth]{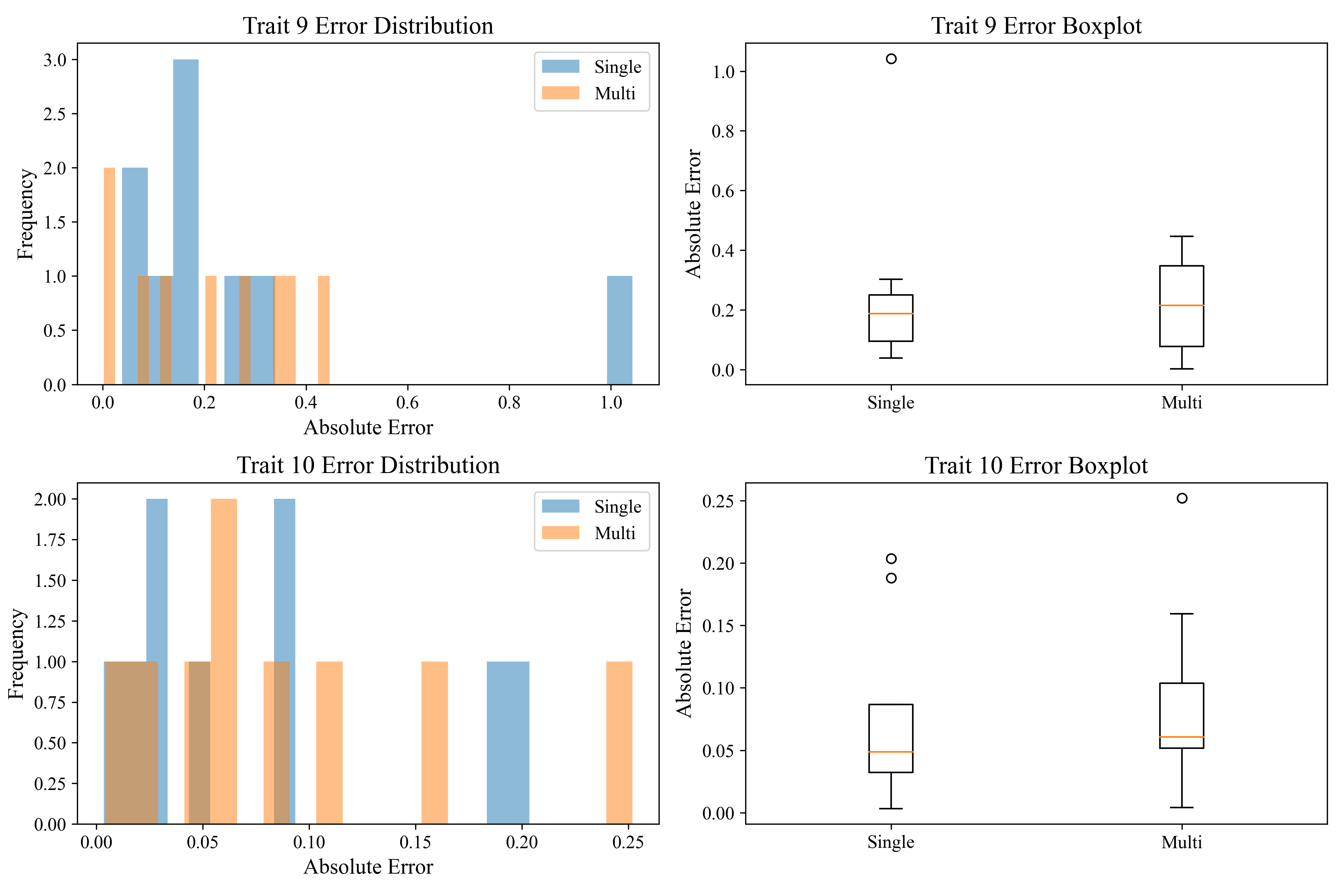}
    \Description{Histograms and boxplots illustrating absolute error distributions for single-task vs. multi-task models.}
    \caption{Error Distribution Comparison Part5. Histograms (left) and boxplots (right) illustrating absolute error distributions for each trait under single-task (blue) vs. multi-task (orange) training.}
    \label{fig:error_distribution_part5.png}
\end{figure}

\subsection{Correlation Analysis}
Table~\ref{tab:pearson_single} presents the single-task Pearson correlations (prediction vs.\ actual), whereas Table~\ref{tab:pearson_multi} shows the same metric for the multi-task model.\footnote{All correlations are computed on the test set.} We also compute the inter-trait correlation matrix (Table~\ref{tab:trait_corr_matrix}, partial snippet below) to highlight which traits exhibit high or low inherent relationships. Notably, Trait 1 and Trait 6 appear moderately correlated in the data, consistent with observing a positive transfer effect in multi-task training for those traits. By contrast, Trait 2’s negative correlation with Trait 1 (about $-0.40$) might explain why its performance degrades in the multi-task model.

\begin{table}[h!]
\centering
\begin{tabular}{l c}
\hline
Trait & Corr(Single) \\
\hline
Trait1 & 0.1067 \\
Trait2 & 0.2750 \\
Trait3 & 0.7406 \\
Trait4 & 0.7720 \\
Trait5 & -0.0056 \\
Trait6 & 0.6092 \\
Trait7 & 0.0728 \\
Trait8 & 0.7056 \\
Trait9 & 0.0385 \\
Trait10 & 0.4416 \\
\hline
\end{tabular}
\caption{Single-task Pearson correlations (test set).}
\label{tab:pearson_single}
\end{table}

\begin{table}[h!]
\centering
\begin{tabular}{l c}
\hline
Trait & Corr(Multi) \\
\hline
Trait1 & 0.6256 \\
Trait2 & -0.3322 \\
Trait3 & 0.4557 \\
Trait4 & -0.0516 \\
Trait5 & 0.1526 \\
Trait6 & 0.5940 \\
Trait7 & -0.1316 \\
Trait8 & 0.7732 \\
Trait9 & 0.4313 \\
Trait10 & 0.3289 \\
\hline
\end{tabular}
\caption{Multi-task Pearson correlations (test set).}
\label{tab:pearson_multi}
\end{table}

\begin{table}[h!]
\centering
\begin{tabular}{lcccccccccc}
\hline
     & T1 & T2 & T3 & T4 & T5 & T6 & T7 & T8 & T9 & T10 \\
\hline
T1   & 1.00 & -0.40 & -0.28 & ... & ... &  0.66 & ... &  0.60 & 0.29 & 0.30\\
T2   & -0.40 &  1.00 & -0.11 & ... & ... & -0.63 & ... & -0.23 & -0.62 & -0.52\\
T3   & -0.28 & -0.11 & 1.00  & ... & ... & -0.25 & ... & -0.61 &  0.04 & 0.52\\
...  &  ... &  ... &  ...  & ... & ... &  ...  & ... &  ...  & ...  & ...\\
T10  &  0.30 & -0.52 & 0.52 & ... & ... &  0.56 & ... & -0.07 & 0.71 & 1.00\\
\hline
\end{tabular}
\caption{Excerpt of inter-trait correlation matrix (test set).}
\label{tab:trait_corr_matrix}
\end{table}

Overall, these correlation patterns align with our earlier discussion of positive vs.\ negative transfer. Traits that share moderate to high correlation typically benefit from multi-task training, whereas tasks that are negatively or weakly correlated may degrade each other’s performance when forced to share capacity.

\subsection{Fine-Tuning and Plasticity Retention}
Finally, we investigate whether \emph{fine-tuning} can restore single-trait performance for the multi-task model. We take the trained multi-task network and fine-tune each trait separately for 10 epochs, only calculating the loss on that specific trait’s output. Table~\ref{tab:fine_tune_mse} shows the MSE on the test set before vs.\ after fine-tuning. 

\begin{table}[h!]
\centering
\begin{tabular}{lcc}
\hline
Trait & MSE(Before) & MSE(After)\\\hline
Trait1 & 75.9286 & 103.2831\\
Trait2 & 158.4957 & 220.2012\\
Trait3 & 29.3389 & 28.9299\\
Trait4 & 52.9163 & 56.4575\\
Trait5 & 58.7907 & 88.0614\\
Trait6 & 3.5989 & 7.2997\\
Trait7 & 6.4702 & 7.2671\\
Trait8 & 0.2756 & 0.2627\\
Trait9 & 0.0667 & 0.0578\\
Trait10 & 0.0054 & 0.0069\\
\hline
\end{tabular}
\caption{Fine-tuning MSE before/after.}
\label{tab:fine_tune_mse}
\end{table}

In most cases (e.g., Trait 3, Trait 8, Trait 9), the fine-tuned error is \emph{similar to or smaller than} the “before” value, indicating successful specialization that recovers single-task-like performance. However, some traits (1, 2, 5) exhibit higher MSE post-fine-tuning, likely due to overfitting on limited training data or difficulty escaping the parameter constraints established by multi-task training. This highlights that while the multi-task model retains a certain level of plasticity, simply fine-tuning does not guarantee improvement in every case; careful hyperparameter choices (e.g., smaller learning rate or early stopping) might be required to ensure consistent gains.

\paragraph{Summary of Findings.}  
The single-task approach generally produces stable performance (average NRMSE $\approx0.3784$ on the test set), whereas the multi-task model shows a mixture of improvements and degradations (average NRMSE $\approx0.3574$ on test). Error distribution plots verify where multi-task learning excels (fewer large outliers, presumably from positive transfer) and where it underperforms (negative transfer). Pearson correlations largely corroborate these observations: strongly related traits benefit, unrelated traits do not. Fine-tuning partially recovers or even surpasses single-task performance in some traits, but fails in others due to potential overfitting or shared-parameter interference. Overall, multi-task learning remains a promising strategy for G2P, but task correlation and model design must be carefully managed to realize consistent gains across all traits.
\section{Data Structure and G2P Analysis Pipeline}

\subsection{Data Acquisition and Preprocessing}
The Rice529 dataset, consisting of 529 inbred rice lines and 4046013 single nucleotide polymorphisms (SNPs), serves as a valuable resource for genomic selection (GS) studies in rice. By capturing a wide array of genetic variation, this dataset supports the investigation of genotype-to-phenotype associations, particularly in predicting key agronomic traits like yield. The comprehensive SNP markers within Rice529 allow researchers to improve the precision of phenotypic predictions, thereby advancing rice breeding efforts.\par

\begin{figure}[!ht]    
\centering
\includegraphics[width=0.9\textwidth]{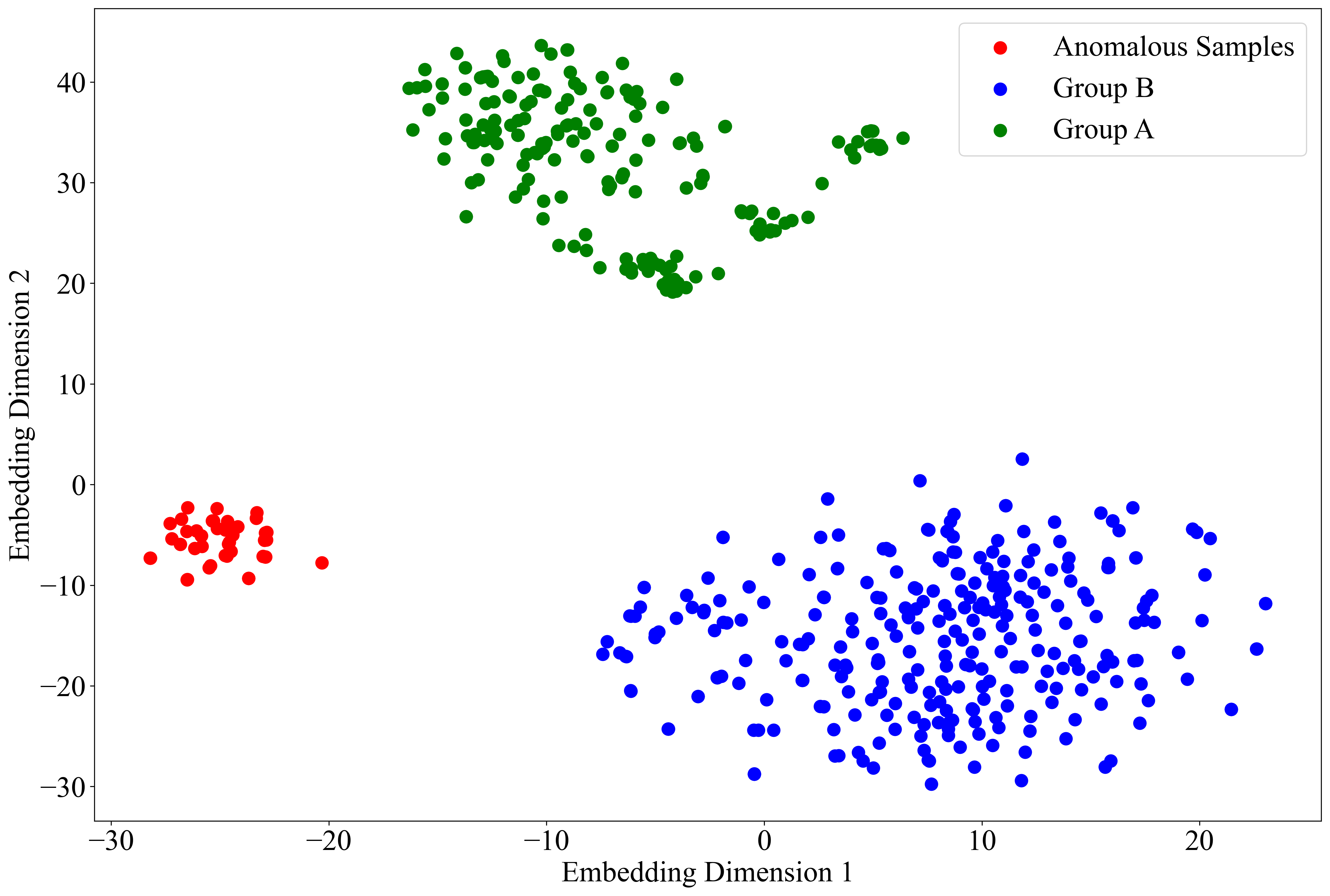}
\caption{t-SNE embedding of the dataset. }
\Description{The t-SNE plot visualizes the genetic diversity among the 529 rice lines, highlighting potential outliers and subpopulation structure within the dataset. The red cluster represents anomalous samples identified as \textit{japonica} or \textit{japonica}-admixed, while the green and blue clusters correspond to the \textit{indica} and \textit{indica}-admixed groups. This visualization aids in understanding the genetic stratification and ensuring the integrity of the genomic selection analysis.}
\label{TSNE_embedding_of_dataset}
\end{figure}

Additionally, the original dataset comprised 4,735 samples with 17,397,026 variants. To align with the phenotypic data, only 529 samples were retained. The majority of the 529 lines were initially considered, based on breeding records, to belong to the indica or indica-admixed subpopulation groups. To identify outliers that belong to the japonica or japonica-admixed groups, t-SNE (t-distributed Stochastic Neighbor Embedding) analysis was performed using 4,046,013 SNPs. Based on this analysis, 44 outliers were identified and excluded (see in Fig. \ref{TSNE_embedding_of_dataset}). After removing these 44 outliers, the resulting t-SNE analysis showed no significant subpopulation stratification within the dataset.\par

\begin{figure}[!ht]    
\centering
\includegraphics[width=0.9\textwidth]{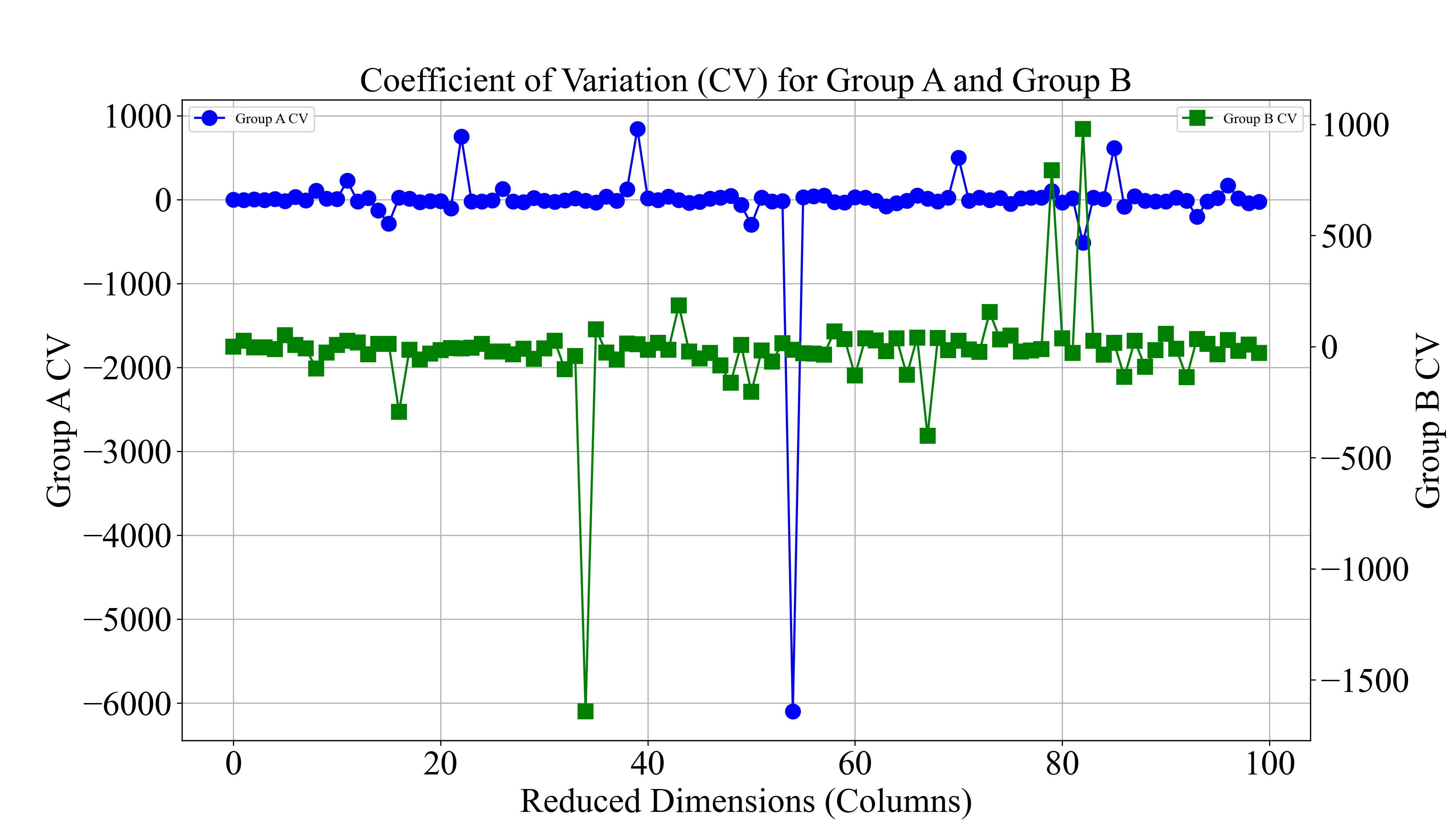}
\caption{Comparison of CV between Group A and Group B across Reduced Dimensions.}
\Description{Group A exhibits relatively stable CV values, while Group B shows greater fluctuations, suggesting differences in stability and potential prediction difficulty between the two groups.}
\label{cv_comparison_group_a_b}
\end{figure}

However, family structure was still considered a potential source of interference in the experimental results. Since including closely related individuals (e.g., full siblings) in both the training and testing sets during model validation can lead to artificially inflated prediction accuracy, it was necessary to control for this family structure. The traditional dataset setup for regression models typically uses the n-fold method, which is suitable for cases where there is strong balance among samples within a population. To determine whether the two rice populations currently under study meet the requirements for the n-fold method, we used the Coefficient of Variation (CV) to measure the balance within the population samples. The measurement results are shown in Fig. \ref{cv_comparison_group_a_b}.\par

\begin{figure*}[!ht]    
\centering
\includegraphics[width=1.0\textwidth]{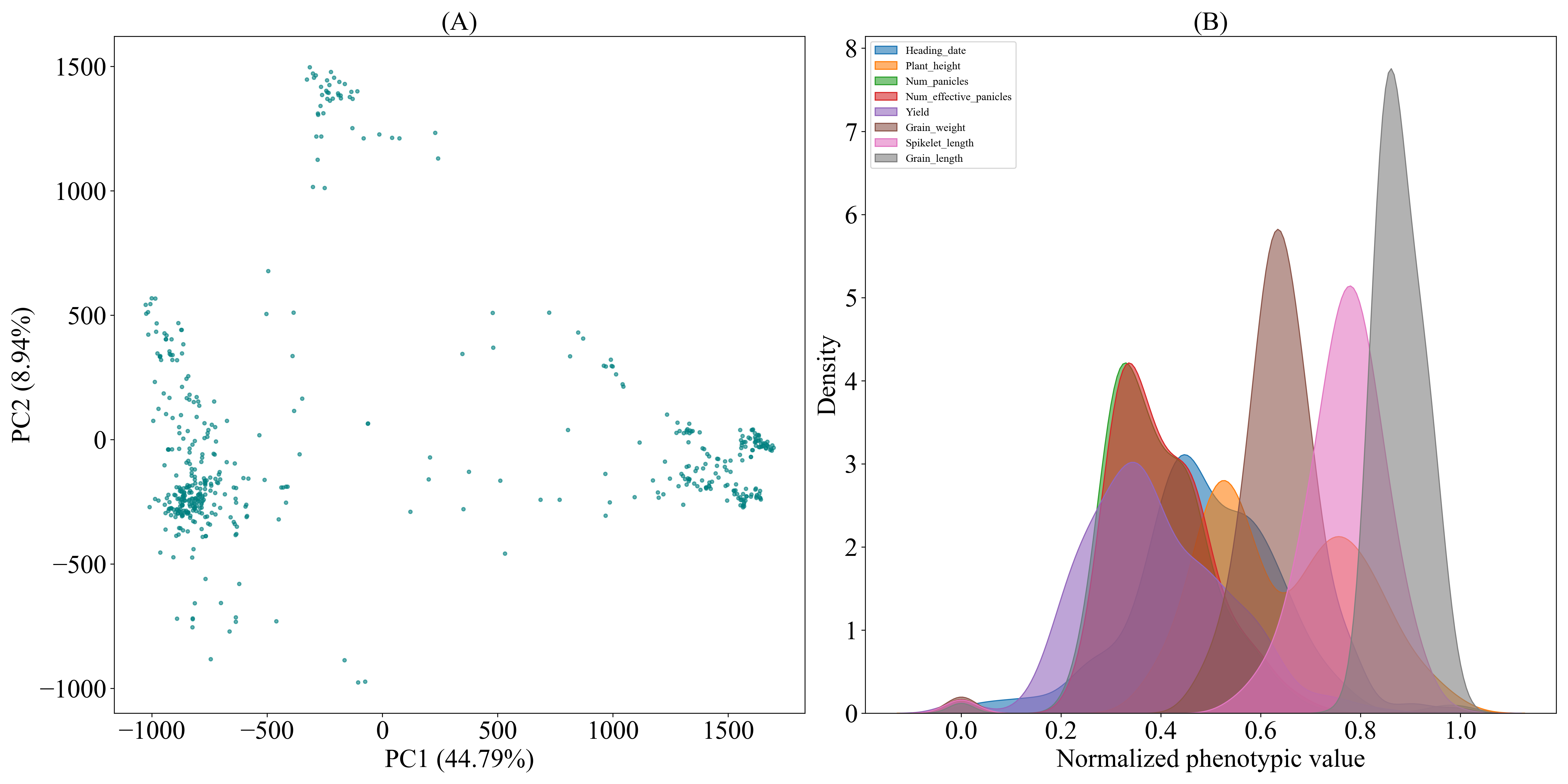}
\caption{Population structures and phenotype distribution of Rice529.}
\Description{Part (A) employs Principal Component Analysis (PCA) to reveal the genetic structure within the population, showing the distribution of Rice529 samples along PC1 and PC2, thus uncovering potential stratification. Part (B) depicts the normalized density distributions of various phenotypic traits within the Rice529 dataset, highlighting the distribution patterns and diversity among different traits. These figures provide crucial visual information for studying the genetic structure and phenotypic diversity in the Rice529 dataset.}
\label{Population_structures_and_phenotype_distribution_of_Rice529}
\end{figure*}

To further illustrate the principal components and distribution within the dataset, we present the Population structures and phenotype distribution of Rice529 dataset(see in Fig. \ref{Population_structures_and_phenotype_distribution_of_Rice529}). Subfigure (A) presents the results of Principal Component Analysis (PCA) for the Rice529 dataset, showing sample distribution along PC1 and PC2, which account for 44.79\% and 8.94\% of the variance, respectively. This visualization reveals clustering patterns and potential population structures, with significant differences in sample distribution across these components. Subfigure (B) displays the normalized distribution of various phenotypic traits using kernel density estimation, highlighting the distribution patterns and variation across traits such as spikelet length and grain length, providing valuable insights for genetic association studies.\par

\subsection{Architecture of the Proposed Method}

In this section, we provide a high-level pipeline for genotype-to-phenotype (G2P) time-series analysis, serving as an overarching structure before detailing our methods. The flow begins by preparing genotype data $\mathbf{G}(t)$ and phenotype data $\mathbf{Y}(t)$, then incorporates a series of preprocessing routines before applying the Learnable Group Transform (LGT). The resultant representations are passed to a classification or regression layer to produce phenotype predictions, trained via cross-entropy loss. Each subsequent subsection offers in-depth explanations of these components, including objective definition, data modeling, and the theoretical underpinnings of LGT. LGT is the core feature extraction module of this paper, which is employed to extract spatial relationships and patterns from the genotype features. It operates by applying a mother filter to the group-transformed input data, which is then processed via convolutional layers to obtain rich, high-level feature embeddings.\par

\begin{figure}[!ht]    
\centering
\includegraphics[width=0.9\textwidth]{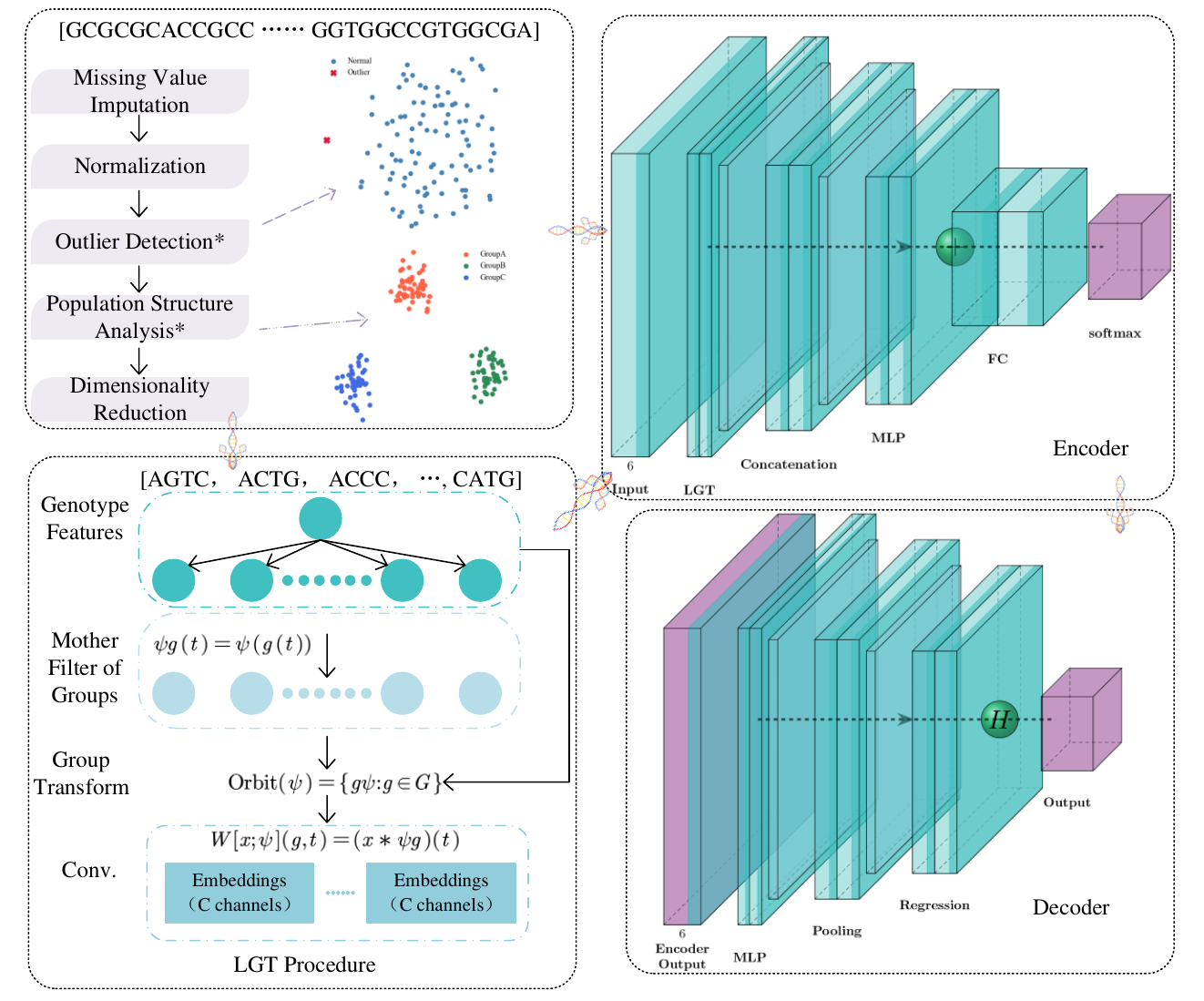}
\caption{Overview of the Proposed Pipeline with LGT, Encoder, and Decoder Architecture.}
\Description{The Encoder module, shown on the right side, takes these transformed features and performs further processing using multi-layered neural network structures such as MLP, concatenation, and softmax layers. This structure allows the model to effectively map the genotype data to a latent representation suitable for phenotype prediction. The Decoder module is responsible for taking the encoded representation and generating the final phenotype prediction. }
\label{FlowChartGroup}
\end{figure}

Overall, this pipeline captures genotype-driven temporal patterns and translates them into phenotype predictions through a combination of carefully chosen transformations and modular optimization steps. Below, we illustrate the main procedure in a flow diagram (See in Fig. \ref{FlowChartGroup}), highlighting where each module (preprocessing, LGT, and prediction) is invoked. The detailed mathematical formulations and implementation strategies follow in the subsequent subsections.\par

\subsection{Objective Function Definition}
We formulate the genotype-to-phenotype (\textbf{G2P}) time series prediction as a sequence classification problem. Let there be:
\begin{itemize}
    \item $N$ samples (indexed by $i = 1,\dots, N$).
    \item $M$ genotype features (e.g., markers or genes) for each sample.
    \item $T$ discrete time points (indexed by $t = 1,\dots, T$).
    \item $Q$ phenotype classes (or phenotype outputs) in a multi-class setting.
\end{itemize}

For each sample $i$ at time $t$, the \textbf{genotype} is denoted by a vector $\mathbf{x}_i(t)\in\mathbb{R}^M$, and the corresponding \textbf{phenotype} is represented by a vector $\mathbf{y}_i(t)\in\{0,1\}^Q$ (one-hot or multi-hot encoding if it is classification). Our model predicts a probability distribution $\widehat{\mathbf{y}}_i(t)\in[0,1]^Q$. 

\paragraph{Loss Function.}
We minimize the cross-entropy loss over all samples and time points:
\begin{equation}
\label{eq:cross_entropy}
\min_{\widehat{\mathbf{y}}}\;-\sum_{t=1}^{T}\sum_{i=1}^{N} \mathbf{y}_i(t)^\top \,\ln \bigl(\widehat{\mathbf{y}}_i(t)\bigr),
\end{equation}
where $\ln(\cdot)$ is applied elementwise. In words, for each $(i,t)$, we measure the negative log-likelihood of the true phenotype class $\mathbf{y}_i(t)$ under the predicted distribution $\widehat{\mathbf{y}}_i(t)$, and then sum over all $i$ and $t$.

\subsection{Data Structures}
We use two key data structures to represent the genotype information and the phenotype outputs (both measured over time):

\begin{enumerate}
    \item \textbf{Genotype Data Matrix} $\mathbf{G}(t)\in\mathbb{R}^{N\times M}$: Each row corresponds to a sample ($N$ total), and each column corresponds to one genotype feature ($M$ total). In many real scenarios, $N \ll M$ because the number of measured genetic markers can exceed the sample size.
    
    \item \textbf{Phenotype Data Matrix} $\mathbf{Y}(t)\in\mathbb{R}^{N\times Q}$: Each row corresponds to a sample and each column to one phenotype dimension ($Q$ total). Often $Q \ll N$, e.g., we may have up to 20 traits or labels to predict for each sample.
\end{enumerate}

These structures allow the model to incorporate both the genotype feature space (which can be large) and the phenotype space (which is typically smaller) at each time step $t$.

\subsection{Mathematical Definition of LGT (Learnable Group Transform)}
The \textbf{Learnable Group Transform (LGT)} constructs a bank of filters by applying a learnable transformation $g$ (an element of a group $G$) to a mother filter $\bm{\psi}$. We denote the mother filter by $\bm{\psi}:\mathbb{R}\to\mathbb{R}$ (or higher-dimensional if needed), and each transformed filter is $\bm{\psi}_g(\cdot)$ for $g\in G$. 

Formally:
\begin{equation}
\label{eq:mother_filter}
\bm{\psi}_g(t) \;=\; \bm{\psi}\bigl(g(t)\bigr)\;, \quad g\in G.
\end{equation}
For a time series genotype signal $\mathbf{x}(t)$, the transform output is defined via convolution with each group-transformed filter:
\begin{equation}
\label{eq:lgt_transform}
W[\mathbf{x};\bm{\psi}](g, t) \;=\; \bigl(\mathbf{x}\ast \bm{\psi}_g\bigr)(t),
\end{equation}
where $(\mathbf{x}\ast \bm{\psi}_g)(t)=\sum_{\tau}\mathbf{x}(\tau)\,\bm{\psi}_g(t-\tau)$. In practice, $\mathbf{x}(t)$ can be multi-dimensional (e.g., $M$ genotype features); thus one can either apply separate filters to each dimension or combine them with channel-wise transformations.

\subsection{Permutation Group Embedding}
To enhance equivariance in a genotype feature network, we choose the permutation group $S_{M}$ (on $M$ elements) as the transformation group for LGT.\footnote{If the model acts on $M$ genotype features, then $S_{M}$ is the set of all permutations among those features.} The idea is to make the learned filters \emph{equivariant} to reindexing (permuting) the genotype dimension.

Specifically, we want for any permutation $\sigma\in S_M$:
\begin{equation}
\label{eq:permutation_equiv}
W[\sigma(\mathbf{x});\bm{\psi}](g, t)
\;=\;
W[\mathbf{x};\bm{\psi}]\bigl(\sigma^{-1}g,\, t\bigr),
\end{equation}
where $\sigma(\mathbf{x})$ permutes the feature indices of $\mathbf{x}$. Implementing this requires a consistent way of permuting the filters (indexed by $g$) so that applying $\sigma$ to the input is equivalent to permuting $g$ by $\sigma^{-1}$ in the filter bank. A permutation operator $P_\sigma$ is introduced to systematically handle how the filter’s input/output channels are permuted.

\subsection{Max--Min Alternating Optimization Strategy}
To train the LGT model efficiently, we adopt a \textbf{max--min alternating} strategy that updates the filter parameters and the transformation parameters in separate steps:

\begin{enumerate}
    \item \textbf{Optimize transformation parameters} $g$: 
    Fix the mother filter $\bm{\psi}$ (and all derived filters). Update the group transformations $\{g\in G\}$ in order to maximize model performance (or equivalently, minimize the loss with respect to $g$). 

    \item \textbf{Optimize filter parameters} $\bm{\psi}$: 
    Fix the transformation parameters $g$ and update $\bm{\psi}$ (the mother filter and its group-based variations) to minimize the loss.

    \item \textbf{Alternate} until convergence: 
    Repeat the above two steps until we reach a stable solution. This yields better performance than naively optimizing all parameters at once, and it helps reduce computational complexity in high-dimensional group transformations.
\end{enumerate}

\subsection{Training Procedure (Pseudocode)}
Because you requested pseudocode (not Python), below is a minimal, language-agnostic flow illustrating the alternating optimization. We denote \texttt{FilterOpt} as the optimizer for $\bm{\psi}$, and \texttt{TransformOpt} as the optimizer for the group transformations $g$.

\begin{algorithm}[!ht]
\caption{Max--Min Alternating Training of LGT}
\begin{algorithmic}[1]
\REQUIRE Training set $\mathcal{D}$ containing genotype--phenotype pairs $\bigl(\mathbf{x}_i(t), \mathbf{y}_i(t)\bigr)$ for $i=1,\dots,N$ and $t=1,\dots,T$. 
\REQUIRE Initial filter parameters $\bm{\psi}^{(0)}$, initial transformation parameters $g^{(0)}$.

\FOR{epoch = 1 to \texttt{maxEpochs}}
    \FOR{each mini-batch $B \subset \mathcal{D}$}
        \STATE \textbf{(A) Update Filter Parameters $\bm{\psi}$:}
        \STATE   Freeze $g$; compute loss $\mathcal{L}$ \textit{w.r.t.} $\bm{\psi}$ using Eq.~\eqref{eq:cross_entropy} plus any regularization.
        \STATE   $\bm{\psi} \leftarrow \bm{\psi} - \texttt{FilterOpt}(\nabla_{\bm{\psi}}\mathcal{L})$ 

        \STATE \textbf{(B) Update Transformation Parameters $g$:}
        \STATE   Freeze $\bm{\psi}$; compute loss $\mathcal{L}$ \textit{w.r.t.} $g$.
        \STATE   $g \leftarrow g - \texttt{TransformOpt}(\nabla_{g}\mathcal{L})$
    \ENDFOR
    \STATE Check for convergence or stopping criteria.
\ENDFOR
\RETURN $\bm{\psi}, g$ \quad (Trained model parameters)
\end{algorithmic}
\end{algorithm}

In each iteration, the filter parameters $\bm{\psi}$ are refined while keeping the transformation $g$ fixed, and vice versa. This implements the max--min objective in an alternating fashion.

\subsection{Group Theory Extension Theorem}
Finally, we generalize LGT to a permutation-equivariant filter bank under action of $S_{M}$. Below is a more formal statement ensuring that permuting genotype features in the input yields a predictable permutation of the LGT output.

\paragraph{Theorem (Permutation Equivariance).}
Let $S_{M}$ be the permutation group on $M$ genotype features, and suppose the LGT filters are constructed so that 
\[
\bm{\psi}_g \;=\; \bm{\psi}_{\sigma g}
\quad
\forall \,\sigma \in S_{M},
\]
meaning the filter for transformation $g$ is identical to that for $\sigma g$. Then the transform $W[\mathbf{x};\bm{\psi}](g, t)$ is equivariant under $S_{M}$. Concretely, 
\[
W[\sigma(\mathbf{x});\bm{\psi}](g,t)
\;=\;
W[\mathbf{x};\bm{\psi}](\sigma^{-1} g,\; t),
\quad
\forall\,\sigma\in S_{M}.
\]
\textit{Sketch of Proof.} By definition, 
\[
W[\sigma(\mathbf{x});\bm{\psi}](g,t) 
\;=\; 
\bigl(\sigma(\mathbf{x})\ast \bm{\psi}_g\bigr)(t).
\]
Since $\bm{\psi}_g = \bm{\psi}_{\sigma g}$ (by assumption), one may rewrite the convolution using the filter indexed by $\sigma g$. In effect, applying $\sigma$ to the input is equivalent to shifting $g$ to $\sigma^{-1}g$ in the filter index, yielding
\[
(\mathbf{x}\ast \bm{\psi}_{\,\sigma^{-1} g})(t)
\;=\;
W[\mathbf{x}; \bm{\psi}]\bigl(\sigma^{-1}g,\,t\bigr).
\]
Hence, $\sigma$ in the input domain leads to the corresponding $\sigma^{-1}$ reindexing in the filter bank, ensuring permutation equivariance. A full detailed proof can be found in extended group-theoretic treatments of LGT.

\subsection*{Remarks on Matrix Dimensions}
\begin{itemize}
  \item \textbf{Genotype Matrix Dimensions.} In practice, $\mathbf{G}(t)\in \mathbb{R}^{N\times M}$ (not $N\times N$). Since $N$ is the number of samples and $M$ is often the (much larger) number of genotype features, adjacency-like square matrices are often not feasible in real data scenarios.
  \item \textbf{Phenotype Matrix Dimensions.} The predicted phenotype matrix $\mathbf{Y}(t)\in\mathbb{R}^{N\times Q}$, where $Q$ (often $\le 20$) is far smaller than $N$. 
\end{itemize}

These clarifications reflect practical genomic data, where each sample has a large genotype feature vector but only a few measured phenotypic traits.

\section{Conclusion}
In this paper, we presented the Learnable Group Transform (LGT) framework, a novel approach for genotype-to-phenotype (G2P) prediction that unifies linear genetic modeling with graph-based and transformer-based deep learning. Our motivation stemmed from the intricate challenges posed by rice genomic data—namely, small sample sizes, high-dimensional markers, and significant population stratification. By leveraging a group-theoretic perspective, LGT imposes structured constraints on the learned filters, enabling it to capture both additive and non-additive (epistatic) interactions while mitigating the risks of overfitting. Through a max–min alternating optimization scheme, we further control model complexity and systematically incorporate diverse transformations of a shared “mother filter.” This design helps reduce the data requirements that typically hamper deep learning methods in breeding contexts.\par

We evaluated our approach on the Rice529 dataset—covering 529 rice accessions that exhibit considerable genetic variation—and compared single-trait vs. multi-trait learning scenarios. In single-trait predictions, LGT outperformed multiple baselines such as CNN, LSTM, and MLP in terms of MSE, RMSE, and correlation metrics. Under multi-task training, we observed both positive transfer (for correlated traits) and challenges of negative transfer (for more divergent phenotypes), underscoring the importance of task-specific adjustments. Nevertheless, LGT’s flexible architecture allowed fine-tuning to partially recover or surpass single-trait performance in many cases, reflecting its capacity to adapt to complex biological relationships among traits. These empirical results suggest that LGT not only captures high-order gene–gene interactions more effectively than conventional methods but also makes better use of limited data by injecting domain-specific inductive biases—such as genome structure, known gene linkages, and population subgroups.\par

Looking ahead, several directions remain open for future work. First, addressing negative transfer via advanced task-weighting schemes or trait-specific decoder heads may further enhance multi-trait training. Second, expanding the model to incorporate environmental variables or gene–environment interactions could yield more robust performance in field trials. Third, extending LGT to an even broader set of crops and larger-scale datasets may test its scalability and generality in real-world genomic selection pipelines. Ultimately, our findings highlight the promise of group-transform-based deep architectures for overcoming the “small numbers of samples, large dimensions of features” problem in modern plant breeding, paving the way for more accurate and efficient identification of elite genotypes.\par

\bibliographystyle{ACM-Reference-Format}
\bibliography{MSample}










\end{document}